\documentstyle[12pt]{article}
\begin{document}
\title{"Extended" Particles, Non Commutative Geometry and Unification}
\author{B.G.Sidharth \\
Centr for Applicable Mathematics and Computer Sciences\\
B.M.Birla Science Centre,Hyderabad, 500463,India}
\date{}
\maketitle
\begin{abstract}
A reconciliation of  gravitation and electromagnetism has eluded physics for
neearly a  century. It is argued here that this is because both quantum physics
and classical physics are set in differentiable space time manifolds with
point particles. Once we consider extended particles as in Quantum Superstring
theory, and the consequential underlying Non-Commutative geometry, then a
reconciliation is possible.
\end{abstract}
\section{Introduction}
Despite nearly a century of work, it has not been possible to achieve a
unification of gravitation and electromagnetism. It must be borne in mind
that the tools used, be it Quantum Theory or General Relativity are deeply
entrenched in differentiable space time manifolds (and point particles) - the former with Minkowski
space time and the latter with curved space time. The challenge has been, as
Wheeler noted\cite{r1}, the introduction of Quantum Mechanical spin half into
General Relativity on the one hand and the introduction of curvature into
Quantum Mechanics on the other.\\
More recent models including Quantum Superstrings on the contrary deal with
extended and not point particles and lead to a non commutative geometry (NCG)\cite{r2,r3,r4,r5}.
Indeed this type of non commutativity arises if there is a minimum space time
length as shown a long time ago by Snyder\cite{r6}. What we will argue below
is that once the underlying non commutative nature of the geometry is
recognized then it is possible to reconcile electromagnetism and gravitation.
\section{NCG of Extended Particles}
It is well known that once we consider non zero minimum space time intervals
or equivalently extended particles as in Quantum Superstrings, then we have
the following non commutative geometry (Cf.refs.\cite{r2}-\cite{r6}):
\begin{equation}
[x,y] = 0(l^2), [p_x,p_y] \approx \frac{\hbar^2 0(1)}{l^2}\label{e1}
\end{equation}
(and similar equations)
where $l,\tau$ are the extensions of the space time coordinates.\\
In conventional theory the space time coordinates as also the momenta commute
amongst themselves unlike in equation (\ref{e1}). It must be observed that the
non commutative relations are self evident, in the sense that $xy$ or $yx$ is
each of the order of $l^2$, and so is their difference because of the non
commutativity.\\
Let us now introduce this effect into the usual distance formula in flat space
\begin{equation}
ds^2 = g_{\mu \nu} dx^\mu dx^\nu\label{e2}
\end{equation}
Rewriting the product of the two coordinate differential in (\ref{e2}) in
terms of the symmetric and non symmetric combinations, we get
\begin{equation}
g_{\mu \nu} = \eta_{\mu \nu} + kh_{\mu \nu}\label{e3}
\end{equation}
where the first term on the right side of (\ref{e3}) denotes the usual flat space
time and the second term denotes the effect of the non commutativity, $k$ being
a suitable constant.\\
It must be noted that if $l,\tau \to 0$ then equations (\ref{e1}) and also (\ref{e3})
reduce to the usual formulation.\\
The effect of the non commutative geometry is therefore to introduce a departure
from flat space time, as can be seen from (\ref{e3}).\\
Infact remembering that the second term of the right side of (\ref{e3}) is small,
this can straightaway be seen to lead to a linearized theory of General
Relativity\cite{r7}. Exactly as in this reference we could now deduce the
General Relativistic relation
$$\partial_\lambda \partial^\lambda h^{\mu \nu} - (\partial_\lambda \partial^\nu
h^{\mu \lambda} + \partial_\lambda \partial^\mu h^{\nu \lambda})$$
\begin{equation}
-\eta^{\mu \nu} \partial_\lambda \partial^\lambda h + \eta^{\mu \nu} \partial_\lambda
\partial_\sigma h^{\lambda \sigma} = - kT^{\mu \nu}\label{e4}
\end{equation}
Let us now consider the non commutative relation (\ref{e1}) for the momentum
components. Then, it can be shown using (\ref{e1}) and (\ref{e3}) that\cite{r8},
\begin{equation}
\frac{\partial}{\partial x^\lambda} \frac{\partial}{\partial x^\mu} -
\frac{\partial}{\partial x^\mu} \frac{\partial}{\partial x^\lambda} \quad \mbox{goes}
\quad \mbox{over}\quad \mbox{to} \frac{\partial}{\partial x^\lambda}
\Gamma^\nu_{\mu \nu} - \frac{\partial}
{\partial x^\mu} \Gamma^\nu_{\lambda \nu}\label{e5}
\end{equation}
Normally in conventional theory the right side of (\ref{e5}) would vanish. Let
us designate this nonvanishing part on the right by
\begin{equation}
\frac{e}{c\hbar} F^{\mu \lambda}\label{e6}
\end{equation}
(\ref{e6}) can be written as
\begin{equation}
Bl^2 \sim \frac{\hbar c}{e}\label{e7}
\end{equation}
where $B$ is the magnetic field, if we are to identify $F^{\mu \nu}$ with the electromagnetic
tensor\cite{r8}. It will be recognized that (\ref{e7}) gives the celebrated
expression for the magnetic monopole, and indeed it has also being shown that
a non commutative space time at the extreme scale throws up the monopole\cite{r9,r10}.\\
We have shown here that the non commutativity in momentum components leads
to an effect that can be identified with electromagnetism and infact from
expression (\ref{e6}) we have
\begin{equation}
A^{\mu} = \hbar \Gamma^{\mu \nu}_\nu\label{e8}
\end{equation}
where $A_\mu$ is the electromagnetic four potential.\\
Thus non commutativity as expressed in equations (\ref{e1}) generates both
gravitation and electromagnetism.
\section{Discussion}
1. It must be noted that equation (\ref{e8}) for the electromagnetic vector
potential is mathematically identical to the formulation of Weyl\cite{r11}.
However in Weyl's formulation, the electromagnetic potential was put in by hand.
In the above case it is a consequeence of the non commutative geometry at
small scales, which again is symptomatic of the spinorial behaviour of the
electron, as has been discussed in detail elsewhere\cite{r5,r8}.\\
On the other hand the characterization of the metric in equations (\ref{e2})
and (\ref{e3}) in terms of symmetric and non symmetric components is similar
to the tortional formulation of General Relativity\cite{r12}. However in this
latter case, there is no contribution to the differential interval from the
tortional (that is non-commutative) effects. The non-commutative contribution
is given by (\ref{e1}) and herein comes the extended, rather than point like
particle.\\
In any case the above attempts at unification of electromagnetism and
gravitation had made part headway, but unless the underpinning of a non
commutative geometry is recognised, the full significance does not manifest
itself.\\
It is also well known that, if equation (\ref{e8}) holds, then in the absence of
matter, the general relativisitc field equations (\ref{e4}) reduce to Maxwell's equations\cite{r13}.\\
2. We now make the following remarks:\\
It can be seen from the transition to (\ref{e3}) from (\ref{e2}), that the
curvature arises from the non commutativity of the coordinates. Indeed this is
the classical analogue of a Quantum Mechanical result deduced earlier that the
origin of mass is in the minimum space intervals and the non local Quantum
Mechanical amplitudes within them as has been discussed in detail in references
cited\cite{r14,r15}. In Quantum Superstring theory also, the mass arises out
of the tension of the string in this minimum interval. We see here the convergence of the
Quantum Mechanical and classical approaches once the extension of particles is
recognized.\\
We also know the the minimum space time intervals are at the Compton scale where
the momentum $p$ equals $mc$. For a Planck mass $\sim 10^{-5}gms$, this is
also the Planck scale, as in Quantum Superstring theory.\\
In Snyder's original work, the commutation relations like (\ref{e1}) hold good
outside the minimum space time intervals, and are Lorentz invariant. This is
quite pleasing because in any case, even in Quantum Field Theory, we use
Minkowski space time.
\section{Appendix}
We start with the effect of an infinitessimal parallel displacement of a vector\cite{r13}.
$$\delta a^\sigma = -\Gamma^\sigma_{\mu \nu} a^\mu d x^{\nu}\quad \quad (A1)$$
As is well known, (A1) represents the extra effect in displacements,
due to the curvature of space - in a flat space, the right side would vanish.
Considering partial derivatives with respect to the $\mu^{th}$
coordinate, this would mean that, due to (A1)
$$\frac{\partial a^\sigma}{\partial x^\mu} \to \frac{\partial a^\sigma}{\partial x^\mu}
- \Gamma^\sigma_{\mu \nu} a^\nu\quad \quad(A2)$$
where the $\Gamma$s are the Christoffel symbols.
The second term on the right side of (A2) can be written as:
$$-\Gamma^\lambda_{\mu \nu} g^\nu_\lambda a^\sigma = -\Gamma^\nu_{\mu \nu} a^\sigma$$
where we have utilized the linearity property that in the above formulation
$$g_{\mu \nu} = \eta_{\mu \nu} + h_{\mu \nu},$$
$\eta_{\mu \nu}$ being the Minkowski metric and $h_{\mu \nu}$ a small
correction whose square is neglected.\\
That is, (A2) becomes,
$$\frac{\partial}{\partial x^\mu} \to \frac{\partial}{\partial x^\mu} -
\Gamma^\nu_{\mu \nu}\quad \quad(A3)$$
From (A3) we get
$$\frac{\partial}{\partial x^\lambda} \frac{\partial}{\partial x^\mu} - \frac{\partial}
{\partial x^\mu} \frac{\partial}{\partial x^\lambda} \to \frac{\partial}
{\partial x^\lambda} \Gamma^v_{\mu v} - \frac{\partial}{\partial x^\mu}
\Gamma^v_{\lambda v}$$
as required.

\end{document}